\documentclass[12pt]{iopart}

\usepackage{graphicx,color,dsfont}
\usepackage{iopams}

\newcommand{\ket}[1]{| #1 \rangle}

\definecolor{gray}{rgb}{.4,.4,.4}
\definecolor{deepgreen}{rgb}{.1,.6,.3}

\begin{document}

\title{Distant entanglement protected through artificially increased local temperature}

\author{Andre R. R. Carvalho$^1$, Marcelo Fran\c{c}a Santos$^{2,3}$}
\address{$^1$ Australian Centre for Quantum-Atom Optics, Department of Quantum Science, Research School of Physics and Engineering, The Australian National University, ACT 0200, Australia}\address{$^2$ Departamento de F\'isica, Universidade Federal de Minas Gerais, Belo Horizonte, Caixa Postal 702, 30123-970, MG, Brazil}
\address{$^3$ Centre for Quantum Technologies, National University of Singapore, 2 Science Drive 3, 117542 Singapore}
\eads{\mailto{andre.carvalho@anu.edu.au},\mailto{msantos@fisica.ufmg.br}}

\begin{abstract}
In composed quantum systems, the presence of local dissipative channels causes loss of coherence and entanglement at a rate that grows with the temperature of the reservoirs. However, here we show that if temperature is artificially added to the system, entanglement decay can be significantly slowed down or even suppressed conditioned on suitable local monitoring of the reservoirs. We propose a scheme to implement the joint reservoir monitoring applicable in different experimental setups like trapped ions, circuit and cavity QED or quantum dots coupled to nanowires and we analyze its general robustness against detection inefficiencies and non-zero temperature of the natural reservoir. 

\end{abstract}
\pacs{42.50.Dv, 03.65.Yz;  03.67.Pp}

\maketitle

\section{Introduction}

Decoherence is considered the villain in quantum information processing, usually dragging previously entangled systems into separable ones. In the last few decades, different approaches to overcome its effects have been explored including quantum error correction codes\cite{Shor:1995,Calderbank:1996,Ekert:1996,Gottesman:1996}, dynamical decoupling~\cite{Viola:1999}, or encoding the system in decoherence free subspaces~\cite{Zanardi:1997,Lidar:1998}. 

Some other strategies make use of the decoherence process itself to try to protect entangled states. Reservoir engineering techniques, for example, explore the production of tailored made reservoir dynamics that have as steady solution the state to be protected~\cite{Poyatos:1996,Carvalho:2001}. Another option is to use the possibility to access the information that leaks to the environment to simply post-select favorable events, or to actively affect the system, as in the case of feedback methods~\cite{Stockton:2004,Mancini:2005,Mancini:2006,Mancini:2007,Wang:2005,Carvalho:2007}. A common feature in these procedures is the requirement of some sort of non-local interaction to counteract the deleterious effect of decoherence on entanglement. These collective interactions can be easily created when the systems to be entangled are close together but can become a challenge for systems that are far apart.

In the case of distant subsystems, the use of local strategies seems to be the most reasonable option from a practical point of view, although it is clear that any tactics based only on local operations would be useless to create entanglement. However, if an initial entanglement is shared between the distant parties, one could hope that local actions could reduce or even stop entanglement decay.   

Recently, local and non-local monitoring of quantum systems and its description in terms of quantum trajectories have been investigated in the context of entanglement decay~\cite{Nha:2004,Carvalho:2007b,Viviescas:2010,Mascarenhas:2010,Vogelsberger:2010} and local entanglement protection~\cite{Mascarenhas:2010a,Mascarenhas:2010b}. In particular, Vogelsberger and Spehner~\cite{Vogelsberger:2010} observed that the average entanglement could be preserved for systems coupled to local infinite temperature reservoirs under an optimal (in the sense that it maximizes the average entanglement) choice of a quantum trajectory unravelling.

In this paper we present a proposal for the implementation of the above idea to preserve entanglement. First we derive the quantum dynamics conditioned on the measurement outcomes to show that the state evolves stochastically but remains entangled. Moreover, at any given time, the original state can be perfectly recovered as long as the detection signal is recorded with unity efficiency. Second, we show how to locally induce an artificial environment that reproduces the dynamics of decay and excitation reservoirs in order to mimic the infinite temperature condition. Finally, we provide a way to jointly, but locally, monitor both the engineered and the natural reservoirs in terms of quantum jumps or homodyne-like measurements. In this way, we give a physical interpretation in terms of a measurement prescription for the unravelling leading to entanglement preservation. Note that the local character of the proposed scheme allows its application to an arbitrary number of qubits, which makes it useful for different sorts of quantum communication and computation protocols. In the end, we also discuss a specific experimental implementation of the scheme and analyse its robustness against imperfections.

\section{Preserving entanglement through local reservoir measurements}

We consider a composite system consisting of qubits that are weakly coupled to their own local and independent dissipative reservoirs at zero temperature with decay rates $\gamma_{-,\alpha}$. We also consider independently and locally engineered reservoirs that incoherently pump each subsystem with rates $\gamma_{+,\alpha}$. A practical way to produce these artificial environments will be presented in section ~\ref{reserveng}. Under the Born and Markov approximations~\cite{Breuer:2002} the time evolution of the system is then given by the Lindblad~\cite{Lindblad:1976} master equation:
\begin{equation}
\label{eq:decoh_general}
\dot{\rho}=\sum_{i,\alpha} \gamma_{i,\alpha} \mathcal{D}[\sigma_{i,\alpha}] \rho ,
\end{equation}
where $\mathcal{D}[\sigma_{i,\alpha}] \rho = \sigma_{i,\alpha}\rho \sigma_{i,\alpha}^\dagger -1/2 (\sigma_{i,\alpha}^\dagger \sigma_{i,\alpha} \rho +\rho \sigma_{i,\alpha}^\dagger \sigma_{i,\alpha})$ with $i=+,-$, and $\alpha$ indicating the subsystem on which the operators are acting. $\sigma_{-,\alpha}$ is the deexcitation operator for qubit $\alpha$ and $\sigma_{+,\alpha}$ its Hermitian conjugate. Note that in the particular case of $\gamma_+ \le \gamma_-$ these engineered reservoirs simulate local thermal reservoirs of average number of photons $\bar{n}=\frac{\gamma_+}{\gamma_- - \gamma_+}$, and particularly of infinite temperature for $\gamma_- =\gamma_+$~\footnote{The infinite temperature regime is to be understood as the limit where $\bar n \rightarrow \infty$ and $\gamma \rightarrow 0$ such that the rates $\gamma_-$ and $\gamma_+$ are the same and remain finite. In this way the system has equal chance of decaying or being excited.}.

This decoherence dynamics can be unravelled in terms of stochastic quantum trajectories in infinitely many ways, each one corresponding to a different scheme to continuously monitor changes in the reservoirs that act on the qubits~\cite{Carmichael:1993,Molmer:1993,Rigo:1997,Wiseman:2000,Wiseman:2010}. Quantum jumps, for example, describe the discrete clicks on a photodetector while homodyne-like measurements are represented by continuous stochastic equations. In what follows we will show how the local monitoring of \eref{eq:decoh_general} in terms of these kind of trajectories can preserve the entanglement in the system.

\subsection{Dynamics in terms of quantum jumps}
\label{sec:jumps}

The most common unravelling is given by choosing the jump operators as the ones appearing explicitly in \eref{eq:decoh_general}, i.e. $J_{i,\alpha}=\sqrt{\gamma_{i,\alpha} dt} \, \sigma_{i,\alpha}$. 
The jump dynamics combines the action of randomly occurring quantum jumps given by the operators $J_{i,\alpha}$ with periods of no-jump evolution when the system evolves according to the operator $M_{\rm NJ} =1 -1/2 \sum_{i,\alpha} J_{i,\alpha}^{\dagger} J_{i,\alpha}$. After the detection of a jump the system evolves to the state $\rho_{J_{i,\alpha}}(t+dt) = \frac{J_{i,\alpha}\rho(t) J_{i,\alpha}^\dagger}{\rm{Tr}(J_{i,\alpha}^\dagger J_{i,\alpha} \rho(t))}$ where the denominator gives the probability ($p_{i,\alpha}$) of detecting a jump $i$ in the subsystem $\alpha$. If no jump is detected in this interval, and assuming perfect detectors, the system then evolves to  $\rho_{NJ}(t+dt) = \frac{M_{\rm NJ} \rho(t) M_{\rm NJ}}{\textrm{Tr}(M_{\rm NJ}\rho(t) M_{\rm NJ})}$. 
The continuous monitoring of the system then indicates the sequence of operators that have to be applied at each time step determining, in this way, the trajectory of the system conditioned on that specific measurement record. In order to recover the unmonitored evolution of the density matrix given by the master equation, one has to average out all the jump and no-jump possibilities for the qubits: $\rho(t+dt)= p_{\rm NJ} \, \rho_{\rm NJ}(t+dt)+\sum_{i,\alpha} p_{i,\alpha} \rho_{J_{i,\alpha}} (t+dt)$.

To analyse the state dynamics, first note that when $\gamma_{-,\alpha} = \gamma_{+,\alpha}=\gamma_{\alpha}$, the no-jump operator is proportional to the identity,
\begin{eqnarray}
M_{\rm NJ}&=&1- \frac{dt}{2}\sum_{\alpha} \gamma_{\alpha}(\sigma_{-,\alpha} \sigma_{+,\alpha}+ \sigma_{+,\alpha} \sigma_{-,\alpha}) = 1-\sum_{\alpha}\frac{\gamma_{\alpha}}{2} dt .
\end{eqnarray}
In this particular case, $M_{\rm NJ}$ commutes with the jump operators $J_{i,\alpha}$, what means that the moment in which jumps are applied do not influence the final state of the system, only the number and type of jumps. In other words, the un-normalized conditional state of the system at time $t$ for a given trajectory will be given by the sequential application of jumps to the initial state:
\begin{equation}
\rho_c(t) =\left( \prod_s^N J_s \right) \rho(0) \left(\prod_s^N J_s\right)^\dagger,
\end{equation}
where, once again, each $J_s$ corresponds to one of the possible $J_{i,\alpha}$ described above, and $N$ is the number of jumps detected in the particular trajectory. 

Also note that any of these jumps immediately destroys all the entanglement originally present in the system, since it identifies the state of the subsystem whose reservoir ``clicked'. For example, if an spontaneously emitted photon is detected in the reservoir acting on a particular qubit (let's call it $A$), then the jump $J_{-,A}$ projects $A$ into its ground state $|0\rangle$. Therefore this choice of monitoring scheme only allows the preservation of entanglement for trajectories where no jumps are detected, which happen with an exponentially decreasing probability in time. 
 Therefore, the averaged entanglement over all possible trajectories also decays, even though it may still last longer than the entanglement of the unmonitored evolution~\cite{Mascarenhas:2010}.

However, as mentioned before, this choice of jumps is not unique. In fact, new jumps defined as $\tilde J_{k}=\sum_i U_{ki}J_i$, with $U$ (left) unitary, leave the master equation unaltered~\footnote{This unitary is not the most general transformation on the jumps that leave the master equation invariant. The addition of a complex number to the jumps with a corresponding change in the effective Hamiltonian is also possible~\cite{Wiseman:2010} but we omitted it here for simplicity.} while generating a distinct set of trajectories. Observe that the no-jump trajectory remains the same, independent of the unitary transformation $U$, and therefore the conditioned state at time $t$ can still be found by the sequential application of the detected jump operators.

 We will now show that a particular choice of jump operators can in fact preserve the entanglement in every trajectory. If instead of $\sigma_{\pm,\alpha}$ we choose as new jumps the linear combinations
  \numparts
 \begin{eqnarray}
 \label{eq:optimaljumps1}
J_{x,\alpha}=\sqrt{\frac{\gamma_{\alpha} dt}{2}}(\sigma_{-,\alpha} + \sigma_{+,\alpha})=\sqrt{\frac{\gamma_{\alpha} dt}{2}} \sigma_{x,\alpha}, \\
 \label{eq:optimaljumps2}
J_{y,\alpha}=i \sqrt{\frac{\gamma_{\alpha} dt}{2}}(\sigma_{-,\alpha} - \sigma_{+,\alpha})=\sqrt{\frac{\gamma_{\alpha} dt}{2}} \sigma_{y,\alpha},
 \end{eqnarray}
  \endnumparts
the jump operations will now be local and unitary, hence they cannot change the entanglement shared by the qubits. Since the product of Pauli operators is again a Pauli operator or the identity, the conditioned state $\rho_c(t)$ of the system at time $t$ in any given trajectory will be given by 
 \begin{equation}
 \label{eq:recovery}
 \rho_c(t) = \bigotimes_{\alpha} \sigma_{Pauli,\alpha} \, \rho(0) \bigotimes_{\alpha} \sigma_{Pauli,\alpha}, 
\end{equation} 
where each $\sigma_{Pauli,\alpha}=\prod_{s}^N J_{s,\alpha}$ is one of $\{\mathds{1},\sigma_{x,\alpha},\sigma_{y,\alpha},\sigma_{z,\alpha}\}$ and is determined by the sequence of jumps recorded on the respective subsystem $\alpha$. In each trajectory entanglement is preserved because the system randomly jumps from one entangled state to another. However, provided that the parts keep track of their sequence of monitored jumps, the initial state can be recovered by locally reversing the final Pauli operation conditioned on the measurement results. 
Note that the scheme is valid for any number of qubits since it only relies on the fact that the jumps are local unitary operations and the state does not change between jumps.

\subsection{Dynamics in terms of diffusive unravellings}
\label{sec:diffusive}

The same idea can be applied to continuous in time stochastic unravelling. For perfect detection efficiency, the evolution of the system is given by~\cite{Rigo:1997,Wiseman:2000,Wiseman:2010}
\begin{eqnarray}
\label{eq:diffusive}
d \rho_{\vec Y}&=& \sum_{i,\alpha} \gamma_{\alpha} \mathcal{D}[\sigma_{i,\alpha}] \rho_{\vec Y} dt  + \sum_{i,\alpha} \sqrt{\gamma_{\alpha}}  \left[(\sigma_{i,\alpha}-\langle \sigma_{i,\alpha} \rangle)\rho_{\vec Y} \, d \xi_{i,\alpha}^* + h.c.  \right],
\end{eqnarray}
 where $d \xi_{i,\alpha}$ are complex Wiener increments satisfying
 \begin{eqnarray}
 d\xi_{i,\alpha} (t) d\xi_{j,\alpha^{'}}^* (t) &=& dt  \delta_{ij} \delta_{\alpha \alpha^{'}},\\
 d\xi_{i,\alpha} (t) d\xi_{j,\alpha^{'}}^* (t)&=& dt  \, u_{ij,\alpha} \delta_{\alpha \alpha^{'}},
 \end{eqnarray}
with $u$ being a complex symmetric matrix subject to the condition $\|u\|_2 \le 1$ for the matrix two-norm. Note that the $\delta_{\alpha \alpha^{'}}$ indicates that all measurements are local and no correlation between the noises in the different subsystems are allowed in our model. The subscript $\vec Y$ in $\rho$ is to remind that its evolution is conditioned on the measurement record given by the complex currents
 \begin{equation}
 \label{eq:Y}
 Y_{i,\alpha} d t =\sum_j dt \sqrt{\gamma_{\alpha}} \langle \sigma_{i,\alpha} + u_{ij,\alpha} \sigma_{j,\alpha}^{\dagger} \rangle + d\xi_{i,\alpha}.
  \end{equation}
 
This time, the freedom in the choice of unravelling lies in the values of the noise correlations encoded in the matrix $u$. Each different $u$ is associated with generalised homodyne-like measurements and corresponds to a particular way of mixing the signals from the monitored system and the noise introduced by external classical sources. To find the value that maximizes the average entanglement one can derive directly an equation for the entanglement in the system~\cite{Guevara:} as has been done for the case of zero temperature reservoir~\cite{Viviescas:2010,Vogelsberger:2010}. In what follows we will show that the choice $u_{ij,\alpha}=-1$ for $i\ne j$ and $u_{ii,\alpha}=0$ lead to perfect entanglement protection for a diffusive monitoring.

To see that, first note that with this choice of $u$ the complex noises can be written as
\begin{eqnarray}
d\xi_{-,\alpha}=\frac{dW_{1,\alpha} + i dW_{2,\alpha}}{\sqrt{2}},\\
d\xi_{+,\alpha}=\frac{-dW_{1,\alpha} + i dW_{2,\alpha}}{\sqrt{2}},
\end{eqnarray}
with the real noises $dW$ obeying $dW_{i,\alpha} (t) dW_{j,\alpha^{'}} (t) = dt  \delta_{ij} \delta_{\alpha \alpha^{'}}$. Using these relations and $\sigma_{-,\alpha}=(\sigma_{x,\alpha} -i \sigma_{y,\alpha})/2$, $\sigma_{+,\alpha}=(\sigma_{x,\alpha} +i  \sigma_{y,\alpha})/2$ , we can express (\ref{eq:diffusive}) in a more convenient form:
\begin{eqnarray}
\label{eq:diffusivePauli}
d \rho_{\vec Y}&=& \sum_{\alpha} \frac{\gamma_{\alpha}}{2} dt \left(\sigma_{x,\alpha}\rho_{\vec Y}\sigma_{x,\alpha} +\sigma_{y,\alpha}\rho_{\vec Y}\sigma_{y,\alpha}-2\rho_{\vec Y}\right)  \nonumber \\
&-&i\sum_{\alpha}\sqrt{\frac{\gamma_{\alpha}}{2}} \left[(dW_{1,\alpha} \sigma_{y,\alpha}+dW_{2,\alpha} \sigma_{x,\alpha}),\rho_{\vec Y}\right]. 
\end{eqnarray}

Now it is easy to show that this evolution can be mapped into the application of local stochastic Hamiltonians defined as $H_{\alpha} = \sqrt{\gamma_{\alpha}}(dW_{1,\alpha} \sigma_y + dW_{2,\alpha} \sigma_x)/\sqrt{2}$. The conditional state of the system at $t+dt$ can be written as
\begin{eqnarray}
&&\rho_{\vec Y}(t+dt)=e^{-i H_{\alpha}} \rho_{\vec Y}(t)e^{i H_{\alpha}} \nonumber \\
&&= \left(1-iH_{\alpha} - \frac{H_{\alpha}^2}{2}\right) \rho(t) \left(1+i H_{\alpha} - \frac{H_{\alpha}^2}{2}\right),
\end{eqnarray}
where in the second step we expanded the exponentials up to first order in $dt$. Replacing the definition of $H_{\alpha}$ and using the correlation properties of the real noises $dW$ one obtains back (\ref{eq:diffusivePauli}). The evolution of the system therefore is equivalent to the application of a sequence of local Hamiltonians to the system, hence, unitary and local. Once again, the state evolves stochastically from one entangled state to another, now in a diffusive way, and entanglement is preserved for each trajectory. Moreover, because each subsystem has the record of their current signals and the local unitaries are state independent,  whenever each party decide to use the qubits for any application, they just need to locally apply a final unitary matrix to revert the system back to its initial state. 

\section{Engineering and monitoring the environment} 
\label{reserveng}

\begin{figure}[h]
\hspace{3cm}
\includegraphics[width=9cm]{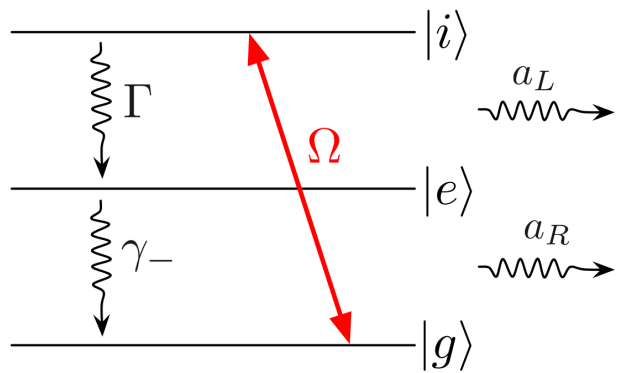}
\caption{Level scheme for engineering the reservoir. The qubit is encoded in levels $\ket{g}$ and $\ket{e}$, which can spontaneously decay back to $\ket{g}$. A classical field couples $\ket{e}$ and $\ket{i}$, which decays back to $\ket{e}$. When the decay rate $\Gamma$ is much larger than the other frequencies of the problem, level $\ket{i}$ can be eliminated and an effective artificial excitation reservoir for the qubit is created.}
\label{fig1}
\end{figure}

We proceed to show the engineering of the artificial temperature reservoir in atomic-like systems. Assume each qubit corresponds to the two lower levels of a three-level atom, $|0\rangle=|g\rangle$ and $|1\rangle=|e\rangle$ and that level $|e\rangle$ spontaneously decays into level $|g\rangle$ at a rate $\gamma_-$, emitting a photon circularly polarized to the right (annihilation operator $a_R$). The detection of this photon corresponds to a $J_- \propto \sigma_{ge}$ jump and upon its detection the state of the atom is certainly level $|g\rangle$. Let us also assume that levels $|g\rangle$ and $|i\rangle$ have the same $m_z$ projection of angular momentum and that an external $\pi$-polarized classical field of intensity $\Omega$ is turned on at the $|g\rangle \rightarrow |i\rangle$ frequency. In this case, if $\Omega \ll \Gamma$ then level $|i\rangle$ can be adiabatically eliminated and the overall effect of turning on the external field is to incoherently pump level $|g\rangle$ back to level $|e\rangle$ at a rate given by $\gamma_+=\frac{4\Omega^2}{\Gamma}$ and with an emission of a photon circularly polarized to the left (annihilation operator $a_L$). Now, the detection of this photon corresponds to a $J_+ \propto \sigma_{eg}$ jump and upon its detection the state of the atom is certainly level $|e\rangle$. Note that if the same scheme is applied on all subsystems and the emitted photons are ignored, the dynamics of the qubits is described exactly by \eref{eq:decoh_general}. Furthermore, if $\frac{4\Omega^2}{\Gamma}=\gamma_-$, then the incoherent pump, together with the spontaneous decay, simulate infinite temperature for the photonic reservoir. 

\begin{figure}[h]
\includegraphics[width=15cm]{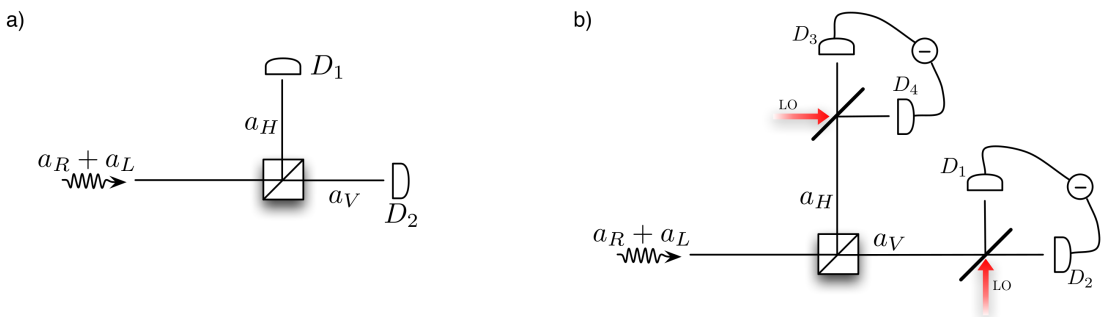}
\caption{Monitoring scheme: photons corresponding to decay and excitation processes are collected and pass through a polarising beam splitter (PBS). For the quantum jump case (a), the detection of photons after the PBS will project the atomic state according to the optimal jump unravelling. For the diffusive case (b), the photons emerging the PBS are combined with local oscillators and the signals recorded providing the currents in \eref{eq:realcurrents1} and \eref{eq:realcurrents2}.} 
\label{fig:monitoring}
\end{figure}

Within this model, it is easy now to describe the setup to implement the entanglement preserving monitoring. For each subsystem the scheme shown in figure \ref{fig:monitoring}a has to be applied. If both photons have the same energy~\footnote{in case $\omega_{ie} - \omega_{eg} = \Delta$ then external fields can be applied to shift the electronic levels to produce photons of the same energy.} and polarized beam splitters (PBS) are placed before the detectors the photons will be combined in linear orthogonal polarizations $a_H=(a_R+a_L)/\sqrt{2}$ and $a_V= i(a_L-a_R)/\sqrt{2}$. Their detection after the PBS will then correspond to the optimal jump operators given in \eref{eq:optimaljumps1} and \eref{eq:optimaljumps2}.

For the diffusive monitoring the scheme is shown in figure \ref{fig:monitoring}b. After the PBS the photons are combined with local oscillators at beam splitters and a homodyne measurement is performed. The currents obtained by subtracting the signals at the pairs of detectors $D_1$ and $D_2$, and $D_3$ and $D_4$ are
\numparts
 \begin{eqnarray}
 \label{eq:realcurrents1}
 I_{1-2}dt= \sqrt{\frac{\gamma}{2}}\langle (a_R -a_L^{\dagger}) - (a_L-a_R^{\dagger})\rangle dt + dW_1, \\
 \label{eq:realcurrents2}
 I_{3-4}dt=-i \sqrt{\frac{\gamma}{2}}\langle (a_R -a_L^{\dagger}) + (a_L-a_R^{\dagger})\rangle dt + dW_2,
 \end{eqnarray}
 \endnumparts
with $dW_i$ real independent increments associated with the noise in the detectors. Now using the fact that the detection of photons $a_R$ and $a_L$ correspond, respectively, to the action of $\sigma_-$ and $\sigma_+$ in the atomic system, one can combine the above signals to obtain the complex currents for the entanglement protecting unravelling, namely, $Y_{-}= I_{1-2}+iI_{3-4}$ and $Y_{+}=-I_{1-2}+i I_{3-4}$. Note that due to the symmetry of the reservoir, when one makes the correspondences  $a_R\rightarrow \sigma_-$ and $a_L\rightarrow \sigma_+$ the average values terms in the signals vanish, indicating that for the protecting monitoring the signals contain only the noise parts. 

From a practical point of view, the difficult part in the above scheme can be the detection of spontaneously emitted photons due to the broad solid angle of emission. For this reason the scheme would be more suited to be implemented in one-dimensional systems where the emitters are embedded in broadband waveguides that allow the propagation of light in only one direction. Such systems are realizable nowadays in different experimental setups such as nanowires~\cite{Chang:2007,Claudon:2010} and superconducting circuit QED~\cite{Houck:2007,Mallet:2009}. 

\section{Analysis of imperfections}
\label{sec:imperfections}

In all the schemes here presented, the biggest problem is the detection inefficiency, specially because the added reservoir increases the emission rates in each qubit, therefore increasing the effect of the decoherence mechanism. We now proceed to calculate the effect of inefficiencies in our proposal. Inefficiency can be treated by including a $\eta$ parameter in the master equation, such that $\eta =1$ means perfect detectors, and by rewriting \eref{eq:decoh_general} as
\begin{equation}
\dot{\rho} = \eta \sum_{i,\alpha} \gamma_{i,\alpha} \mathcal{D}[\sigma_{i,\alpha}] \rho + (1-\eta) \sum_{i,\alpha} \gamma_{i,\alpha} \mathcal{D}[\sigma_{i,\alpha}] \rho.
\end{equation}
The part proportional to $\eta$ corresponds to the one about which we can obtain information, therefore the one we can unravel. The remaining part corresponds to a decoherence evolution, ``slowed' by a factor $(1-\eta)$, due to the information that is flowing to the reservoir and not being retrieved. 

We shall now analyse the effect of detection inefficiencies in the quantum jump description. The most favourable unravelling described in Section~\ref{sec:jumps} has a no-jump proportional to identity and jumps proportional do $\sigma_x$ and $\sigma_y$. The difference now is that one has to add to the no-jump evolution a term corresponding to the decoherence part proportional to $(1-\eta)$. Despite this difference, the analysis remains basically unaltered since the new no-jump evolution still commutes with the jump operations. It suffices to check this for one of the qubits since operations in different subsystems trivially commute. If one writes the local decoherence part in terms of $\sigma_x$ and $\sigma_y$ as in (\ref {eq:diffusivePauli}) one can easily verify that the action of a jump followed by decoherence ($\mathcal{D}[\sigma_x](\sigma_i \rho \sigma_i) +\mathcal{D}[\sigma_y](\sigma_i \rho \sigma_i)$) coincides with the evolution with decoherence followed by the action of a jump ($\sigma_i(\mathcal{D}[\sigma_x]\rho +\mathcal{D}[\sigma_y]\rho)\sigma_i$), where $i$ is either $x$ or $y$.

As a result, the evolution of the density matrix of the system given that $N$ jumps were detected will be given by the mixed state version of \eref{eq:recovery}:
\begin{equation}
\label{eq:recovery_eta}
\rho_c(t) = \bigotimes_{\alpha} \sigma_{Pauli,\alpha} \,  \rho(t)\bigotimes_{\alpha} \sigma_{Pauli,\alpha},
\end{equation}
where $\rho(t)$ is given by the integration of equation $\dot \rho =(1-\eta) \sum_{i,\alpha} \gamma_{i,\alpha} \mathcal{D}[\sigma_{i,\alpha}] \rho$. Note that this equation is equivalent to the unmonitored one except for the final application of local unitaries (that can be reversed since the parties know all the detected jumps), and the factor $(1-\eta)$ corresponding to the detection efficiency. As previously noted, if the detectors are perfect, entanglement if fully preserved. Otherwise, the detection inefficiency becomes simply a multiplier factor on the decoherence rate. 

\begin{figure}[h]
\hspace{3cm}
\includegraphics[width=12cm]{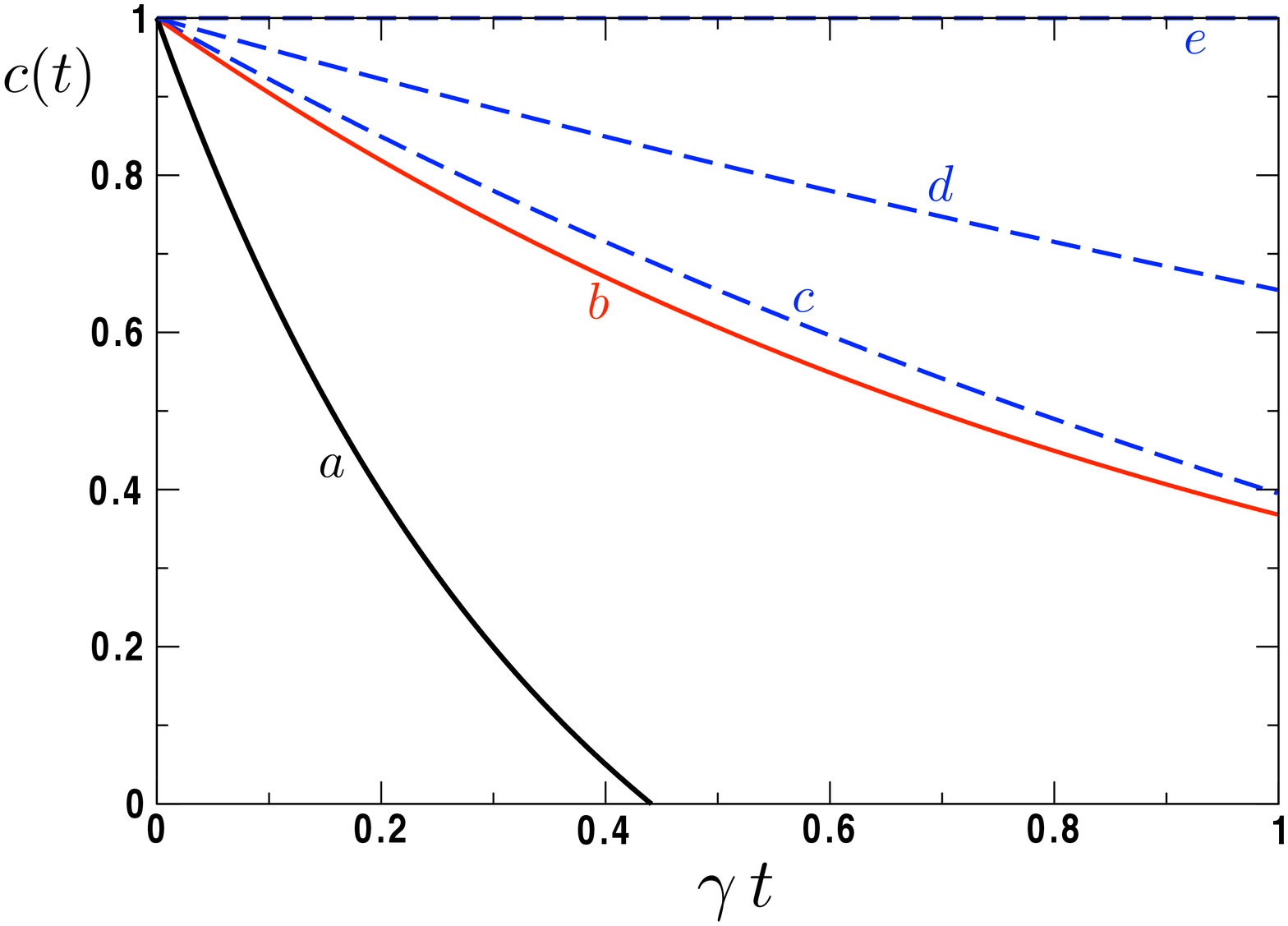}
\caption{Concurrence as a function of time for unmonitored reservoirs at infinite (a) and zero (b) temperatures. Dashed curves represent the average concurrence for the engineered reservoir under monitoring for $\eta=0.8$ (c), $0.9$ (d) and $1.0$ (e).} 
\label{fig:results}
\end{figure}

We are now in position to compare our scheme with both the monitored dynamics without the engineered reservoir and the unmonitored evolution. For that we consider the particular case of 2 qubits, $A$ and $B$, that share initially a maximally entangled state of the form $\ket{\psi_0} =(\ket{01}+\ket{10})/\sqrt{2}$. We also assume that the decay rates are the same ($\gamma_A=\gamma_B=\gamma$). For this state, the unconditional entanglement, as measured by concurrence~\cite{Wootters:1998}, decays as $c(t)=e^{-\gamma t}$ and $c(t)=e^{-2\gamma t}+e^{-4\gamma t}/2 -1/2$ for zero and infinite temperature reservoirs, respectively~\cite{Carvalho:2007a}. The corresponding curves are shown by the solid lines in figure \ref{fig:results}: black (a) for $T_{\infty}$ and red (b) for $T_0$. At zero temperature any local monitoring with $\eta=1$ gives an average entanglement of $e^{-\gamma t}$\cite{Carvalho:2007b, Viviescas:2010}, coinciding with the master equation result (b). The concurrence for the entanglement conditioned on jump monitoring including the engineered reservoir can be calculated using (\ref{eq:recovery_eta}) and gives $c(t)=e^{-2\gamma(1-\eta) t}+e^{-4\gamma (1-\eta) t}/2 -1/2$. The dashed blue curves (c) and (d) represent the cases with $\eta=0.8$ and $\eta=0.9$, respectively, and show that, for the time scale shown in the figure, monitoring preserves entanglement better than not adding the artificial reservoir. For any $\eta<1$, the dynamics will always have an infinite temperature component (with rate $\gamma(1-\eta)$) and the solution~\cite{Carvalho:2007a} will lead to disentanglement of the system at a finite time~\cite{Diosi:2003}, which does not occur for initial Bell states at zero temperature. This means that the curves for the monitored system with artificial reservoir will have to cross the curve of the unmonitored system for the original zero temperature bath. The higher the detection efficiency, the longer our scheme will be advantageous as compared to the unconditional evolution. Obviously, for $\eta=1$ (curve e) one has perfect protection for all times. Although this analysis has been presented for the case of two qubits, we should emphasise that it is also valid for multipartite systems, with perfect protection occurring for ideal monitoring and decline in performance as the detection inefficiency increases.

For a diffusive unravelling, the inefficient evolution is not simple to calculate as in the jump case because now the sequence of evolutions do not commute with the unmonitored evolution. However we can still numerically simulate the equations and the result obtained follows exactly the curves for the jump monitoring. Therefore, since both kinds of monitoring perform equally well theoretically, choosing one monitoring scheme over the other is a matter of finding the one that can be implemented with the best possible detection efficiency. 

Finally, if the reservoirs have natural temperature, then the incoherent pump needs to be weaker in order to balance out both reservoirs. The equation in this case will be similar to the one with inefficiencies, with the natural excitation reservoir, which is not monitored, playing a role similar to that of an additional inefficiency to the detection scheme. 

In conclusion, we showed a way to engineer an artificial reservoir that increases the temperature of the system and yet can be used to decrease the rate of entanglement decay, or even stop it completely, when the system is properly (locally) monitored.  With the use of simple optical elements, we showed how the monitoring scheme that preserves entanglement can be implemented and derived a way to recover the initial state from the measurement record using only local unitary operations. In this way, the loss of information on the system due to the interaction with the environment, which, when irreversible, is ultimately the poison responsible for the decoherence process itself, can also be the vaccine used to fight decoherence as long as the information can be retrieved by a suitable monitoring scheme. 

\ack
We would like to acknowledge the support from Center for Quantum Technologies at the National University of Singapore. ARRC acknowledges financial support by the Australian Research Council Centre of Excellence program and MFS acknowledges the support of the National Research Foundation and the Ministry of Education of Singapore.

\section*{References}
\bibliographystyle{unsrt}
\bibliography{$HOME/ARTICLES/allbib}

\end{document}